\documentclass[aps,prl,preprint,superscriptaddress]{article} 

\usepackage[utf8]{inputenc}
\usepackage[left=3cm,right=3cm,top=3cm,bottom=3cm]{geometry}

\usepackage{amsmath}
\usepackage{amsfonts}
\usepackage{amssymb}
\usepackage{mathtools}
\usepackage{braket}
\usepackage{soul}
\usepackage{cite}

\usepackage{graphicx}

\usepackage{color}
\definecolor{nicered}{rgb}{0.7,0.1,0.1}
\definecolor{nicegreen}{rgb}{0.1,0.5,.1}

\usepackage{subcaption}

\usepackage{stackrel}

\usepackage{multirow}

\usepackage{float}

\usepackage{authblk}

\usepackage{comment}

\usepackage{eso-pic}
\newcommand{\reportnum}[2]{
  \AddToShipoutPictureBG*{%
    \AtPageUpperLeft{%
      \hspace{0.75\paperwidth}%
      \raisebox{#1\baselineskip}{%
        \makebox[0pt][l]{\textnormal{#2}}
  }}}%
}

\usepackage[colorlinks=true
,urlcolor=magenta
,anchorcolor=black
,citecolor=blue
,filecolor=black
,linkcolor=red
,menucolor=black
,linktocpage=true
,pdfproducer=medialab
]{hyperref}

\usepackage[compat=1.1.0]{tikz-feynman}

\usepackage{refcount}

\title{Impact of the $a_1(1260) \pi$ cascade contribution on $D^0 \to \pi^+ \pi^- \ell^+ \ell^-$ decays}

\date{}

\author[a,b]{Eleftheria~Solomonidi}
\author[c]{Luiz~Vale~Silva}

\affil[a]{\it PSI Center for Neutron and Muon Sciences, 5232 Villigen PSI, Switzerland}
\affil[b]{\it Physik-Institut, Universit{\"a}t Z{\"u}rich, Winterthurerstrasse 190, 8057 Z{\"u}rich, Switzerland}
\affil[c]{\it Departamento de Matem\'{a}ticas, F\'{i}sica y Ciencias Tecnol\'{o}gicas,

Universidad Cardenal Herrera-CEU, CEU Universities,

46115 Alfara del Patriarca, Val\`{e}ncia, Spain}

\begin{document}
\reportnum{-6}{ZU-TH 73/25}

\maketitle

\begin{abstract}
For the first time, we incorporate the cascade-type topology $D^0\to \pi^- a_1^+(1260)(\to\pi^+\rho^0(\to\ell^+\ell^-))$ to describe the recently measured rare charm-meson decays. We find that it comprises one of the largest contributions to the decay rate and obtain an unprecedented agreement of the Standard Model prediction with the available LHCb data, while also aptly describing four-body non-leptonic transitions. In presence of this cascade component, we predict that some angular observables acquire non-zero values, further stressing its phenomenological relevance and its potential for constraining new physics.
\end{abstract}

\section{Introduction}

Flavour physics plays a pivotal role in tests of the Standard Model (SM) and indirect searches for New Physics (NP). While the theoretical and experimental programme is at a remarkably advanced level for bottom- and strange-meson processes, the charm counterpart is at an earlier stage of development. However, the investigation of charm-decay phenomena is crucial and unique, as charm is the only decaying heavy up-type quark that is bound in hadrons. In particular, charm processes are excellent candidates for NP effects to be unveiled, as they enjoy a very effective Glashow-Iliopoulos-Maiani (GIM) mechanism because of the lightness, compared to the electroweak scale, of the down-type quarks that appear in the loops, thus suppressing flavour-changing neutral currents.

Currently the most intriguing experimental measurement is the CP violation that has been clearly observed in the difference of hadronic decays $D^0\to\pi^+\pi^-$ and $D^0\to K^+K^-$ \cite{LHCb:2019hro}, while Ref.~\cite{LHCb:2022lry} indicates the dominance of CP violation from $D^0\to\pi^+\pi^-$. The existing theoretical calculations point to a value for the direct CP asymmetry that is much smaller than the experimental one. However, the implemented approaches may require further investigation: in particular, the method of Refs.~\cite{Khodjamirian:2017zdu, Lenz:2023rlq} is based on the framework of light-cone sum rules, for which more tests in the charm sector would be desirable. On the other hand, the data-driven approach of Ref.~\cite{Pich:2023kim} considers hadronic final-state interactions and is based on the assumption that the pion-pair final state rescatters predominantly to a pair of kaons. This assumption can be challenged by an analysis based on the Bethe–Salpeter formalism
in the context of other decay environments \cite{Ropertz:2018stk}, where the channel of four pions appears to be mixing sizably with the two pions at energies close to the mass of the $D$ mesons. While the incorporation of the four-pion channel into the fully data-driven approach is currently unfeasible, it is required
in order to completely understand the dynamics of the two-pion charm-meson decay mode. A description of those weak decays accounting for non-perturbative QCD effects can be achieved by considering a number of intermediate states comprised of strongly decaying resonances. As some of the appearing resonances have been extensively studied in effective theories and models, strong phases can be incorporated through their (data-driven) line shapes. Recent amplitude analyses of the decays to four pions as well as to two pions and two kaons \cite{dArgent:2017gzv, LHCb:2018mzv, BESIII:amplitudean} explore this approach and provide some enlightening results.

The rare decays to light hadrons and two charged leptons have also received attention in the last years. On the experimental front, the recent analyses of Refs.~\cite{LHCb:2017uns, LHCb:2018qsd, LHCb:2020car, LHCb:2021yxk, LHCb:2024hju, LHCb:2024ely, LHCb:2025bfy} have provided an unprecedented volume of information. On the theory side, it is established that the semileptonic operators are very suppressed in the charm decays, namely $C_9$ is about 10 times smaller than the equivalent Wilson coefficient in bottom decays, while $C_{10}$ vanishes at order $G_F\cdot \alpha$ \cite{Fajfer:1998rz,Cappiello:2012vg,deBoer:2016dcg}. Therefore, the decay rate overwhelmingly stems from non-local insertions of four-quark operators, in association with the electromagnetic hamiltonians of quark and lepton currents. Another important consequence of the suppression of the local operators is that a number of angular observables which require a non-zero $C_{10}$ are extremely suppressed
in the SM. While this property is very useful in the search for NP, until a clear signal is experimentally observed the calculation of the long-distance component of those null-test observables, which boosts the effects from NP, remains indispensable for setting meaningful bounds on the NP-driven Wilson coefficients. 

As supported by data, the dilepton pair from rare semileptonic transitions mainly comes from the electromagnetic decay of a vector resonance, a feature that is incorporated in phenomenological works
\cite{Burdman:2001tf, Cappiello:2012vg, DeBoer:2018pdx, Bharucha:2020eup, Fajfer:2023tkp, Bansal:2025hcf}. The calculations of $D^0\to\pi^+\pi^-\ell^+\ell^-$ also model the production of the dihadron pair via the mediation of a vector ($\rho(770)^0\equiv \rho^0$ or $\omega(782)\equiv \omega$) or scalar ($f_0(500) \equiv \sigma$) resonance which decays strongly \cite{Fajfer:2023tkp}. Specifically, they express the decay amplitudes via intermediate quasi-two-body (Q2B) topologies, whereby the charm meson decays weakly to the two intermediate-state resonances that subsequently decay to the final-state particles. An additional normalization factor and a constant phase are assigned to each decay chain so as to encapsulate possible further QCD effects. This approach is the same as the isobar model that is implemented in the experimental amplitude analyses of Refs.~\cite{dArgent:2017gzv, LHCb:2018mzv, BESIII:amplitudean}. Such theoretical calculations within the SM so far succeed at giving an adequate qualitative description of the process when directly compared to the experimental data \cite{Fajfer:2023tkp, Gisbert:2024kob}. However, there are some considerable tensions both in terms of the decay rate distributions as well as the CP-symmetric angular observables. While some of the tensions might be attributed to experimental shortcomings,\footnote{See relevant comments and footnote 12 in Ref.~\cite{Fajfer:2023tkp}.} the most likely explanation behind systematic deviations in the distribution over the dipion mass is the presence of some theoretically unaccounted-for contributions. Additionally, the angular observables that identically vanish in the theoretical model (irrespective of any potential NP) present some non-zero values with a significance of a few standard deviations.

Since the intermediate states of the rare decays also appear as intermediate states of the $4\pi$ or $2\pi 2K$ decays (which are however populated by many additional combinations of resonances, as in this case scalar resonances can also produce efficiently the second hadron pair in lieu of the dilepton), it is instructive to note which decay chains dominate the hadronic decay rates. In both $4\pi$ amplitude analyses \cite{dArgent:2017gzv, BESIII:amplitudean}, the decay chain $D^0\to \pi^- a_1(1260)^+(\to\pi^+\rho^0(\to\pi^+\pi^-))$, which is of the cascade-type topology, where two of the pions are produced consecutively and not at the same vertex, comes out as the largest contributor to the decay rate, with a branching fraction much larger than for instance the chain $D^0\to\rho^0(\to\pi^+\pi^-)\rho^0(\to\pi^+\pi^-)$.\footnote{
The decay mode $D^0\to K^+K^-\ell^+\ell^-$ receives the cascade-type contribution $D^0 \to K^- K_1 (1270)^+$ with the strange axial-vector meson decaying to $\rho^0 (\to \ell^+ \ell^-) K^+$, which is expected to be of some importance based on the amplitude analysis of charm-meson decays to $2\pi 2K$ \cite{LHCb:2018mzv}.}
This result is qualitatively expectable, as in the large-number-of-colors (large-$N_C$) limit the amplitude for the cascade decay is created from an insertion of the Fermi operator $Q_1$, which has a Wilson coefficient about three times larger than the Wilson coefficient of $Q_2$, which appears as the leading contribution in all the Q2B topologies.

Given this fact and since our SM calculation \cite{Fajfer:2023tkp} for the rare decays appears still incomplete, we are motivated to revisit our analysis by introducing, in addition to the
aforementioned 
components, the cascade topology of the $a_1(1260)$, while still keeping to the isobar-like method.
We note that the existence of several axial-vector resonances appearing in a cascade-type topology have been experimentally confirmed in the analogous decays in the bottom system $B^0\to K^+\pi^- \mu^+\mu^-$: most notably $B^0\to K^+ Z(4430)^-(\to \psi' \pi^-)$ \cite{LHCb:2014cascade} as well as $B^0\to K^+ Z(4200)^-(\to J/\psi\, \pi^-)$ \cite{Belle:2014cascade}. In this latter case, however, due to the four-quark content of those exotic resonances the corresponding branching fractions are small.

As per our previous work \cite{Fajfer:2023tkp}, in Section~\ref{seb:implementation} we appropriately assign free normalization factors and constant phases to be fitted to the experimental mass distributions. Interference effects are also taken into account, which stem from the presence of the cascade topology in both the $S$- and $P$-waves of the pion pair. As discussed in Section~\ref{sec:fitting}, based on the values of the free parameters as extracted from the fit, we proceed to estimate how various angular observables are modified in the presence of the new contribution. Finally, in Section~\ref{seb:comparisons} we draw a comparison between our results and the amplitude analyses of the hadronic decays, in an attempt to evaluate the consistency of the resonance-mediated model and the universality of hadronic effects in charm decays.
We conclude in Section~\ref{sec:conclusions}.

\section{Implementation and special features of the cascade contribution}\label{seb:implementation}

For the purposes of the needed precision, due to the features of the short-distance dynamics described in the previous section,
the effective hamiltonian for the decays $c\to u \mu^+\mu^-$ can be reduced to 
\begin{equation} \label{eq:H_eff}
	\mathcal{H}_{\rm eff} = \frac{G_F}{\sqrt{2}} \left[ \, \sum^2_{i = 1} C_i (\mu) \left( \lambda_d Q_i^d + \lambda_s Q_i^s \right)  \right] + \mathrm{h.c.} \,,
\end{equation}

\noindent where $
	\lambda_q = V^\ast_{c q} V_{u q}$, $q = d, s$, and  the operators appearing are the following:

\begin{eqnarray}\label{eq:operator_list}
	 Q_1^q = ( \overline{q} c )_{V-A} ( \overline{u} q )_{V-A} \,, \quad
	&& Q_2^q = ( \overline{q}_j c_i )_{V-A} ( \overline{u}_i q_j )_{V-A}
    \,, \nonumber \quad q = d, s \,,
\end{eqnarray}
where $ (V - A)_\mu = \gamma_\mu (\mathbf{1} - \gamma_5) $, $i, j$ are colour indices, and $\mu \sim \overline{m}_c (\overline{m}_c)$ is the renormalisation scale.

The $S$-matrix elements can be schematically written as follows: 

\begin{eqnarray}\label{eq:Q2B_schematic_S_matrix}
     \langle \pi^+ \pi^- \ell^+ \ell^- | S |D^0 \rangle^{(\mathrm{Q2B})} = && \langle \pi^+ \pi^- \ell^+ \ell^- |  \int d^4 x \, d^4 w \, d^4 y \, d^4 z \\
    && T \{ \mathcal{H}_{em}^{\rm lept}(z) \, \mathcal{H}_{\mathcal{V}\gamma}(y) \, \mathcal{H}_{\mathcal{R} \pi\pi}(w) \, \mathcal{H}_{D \mathcal{R} \mathcal{V}}(x) \} |D^0 \rangle \,, \nonumber
\end{eqnarray}
for the Q2B topologies, where $\mathcal{R}=\sigma, \rho^0$ or a small isospin-violating $\omega$  component, while $\mathcal{V}=\rho^0,\omega,\phi(1020)\equiv \phi$ are the vector resonances which couple to a single photon via the electromagnetic hamiltonian $\mathcal{H}_{\mathcal{V}\gamma}$; and
\begin{eqnarray}
\label{Sforcascade}\label{eq:casc_schematic_S_matrix}
    \langle \pi^+ \pi^- \ell^+ \ell^- | S |D^0 \rangle^{(\mathrm{casc})} = && \langle \pi^+ \pi^- \ell^+ \ell^- |  \int d^4 x \, d^4 w \, d^4 y \, d^4 z \\
    && T \{ \mathcal{H}_{em}^{\rm lept}(z) \, \mathcal{H}_{\mathcal{V}\gamma}(y) \, \mathcal{H}_{\mathcal{A} \mathcal{V}\pi}(w) \, \mathcal{H}_{D \mathcal{A} \pi}(x) \} |D^0 \rangle \,, \nonumber
\end{eqnarray}
for the cascade topologies.
The comparison between Eqs.~\eqref{eq:Q2B_schematic_S_matrix} and \eqref{eq:casc_schematic_S_matrix} contrasts well the two different cases. As we will see later, a crucial difference between them is found in the kinematics.

We only consider the axial-vector resonance $ \mathcal{A} = a_1(1260)\equiv a_1$. Further cascade-type decays are neglected given their reported suppression in the $D\to 4\pi$ decays from amplitude analyses and the lesser known nature of the heavier axial-vector resonances. In contrast, the $a_1$ resonance has been extensively studied in various contexts \cite{Ecker:2002cw,Cheng:2003bn,Gronau:2005kw,Cheng:2010vk,Dalseno:2019kps}. It plays a major role in the hadronic decays of the $\tau$ lepton \cite{GomezDumm:2003ku,Dumm:2009va,Dumm:2009kj} and it decays predominantly through the vector resonance $\rho^0$. Hence, for the purposes of the present work, we consider sufficiently precise to model the decay $a_1\to \pi \ell^+\ell^-$ as the consecutive decays $a_1\to\rho^0\pi$ and $\rho^0\to\ell^+\ell^-$.
We model the decay $a_1\to\rho^0\pi$ with a constant coupling, namely: 
\begin{equation}\label{eq:a1_rho_pi_ME}
    \langle \rho (q,\lambda_\rho) \pi (p_1) |H_{a_1\rho\pi}| a_1 (k,\lambda_{a_1})\rangle = \epsilon_{(a_1)} (k,\lambda_{a_1}) \cdot \epsilon^*_{(\rho)}(q,\lambda_\rho)\,  g_{a_1\rho\pi} \,.
\end{equation}
We follow closely the parameterisation of $a_1$ implemented in the CLEO-c amplitude analysis \cite{dArgent:2017gzv} and use their data-driven lineshape, which is based on Ref.~\cite{Kuhn:1990ad} and with which they find $m_{a_1}=1225\pm 22$ MeV and $\Gamma_{a_1}=430\pm 39$ MeV.

As per our previous work \cite{Fajfer:2023tkp}, the decay amplitudes of the $D$ meson to the intermediate
hadrons are calculated at the large-$N_C$ limit. This can be realised in different topologies; for the Q2B decays these were previously named W-, J- and A-type. For those decays, following the same strategy as before, we do not consider the A topology, corresponding to annihilation diagrams with the photon emitted from the quark legs of the final states and which is qualitatively expected to be suppressed in naive factorization (see, e.g., Ref.~\cite{Bauer:1986bm}). This is well justified as said topology vanishes in the case of $D\to \rho\rho$ and $D\to\rho\omega$ (taking $\rho$ and $\omega$ to be approximately degenerate), violates the Zweig rule in the case of $D\to\rho\phi$ and $D\to\sigma\phi$, and can be absorbed into the W topology constants for $D\to \sigma \rho$ and $D\to\sigma\omega$.\footnote{
Regarding the Q2B contribution,
the suppression of the A topology can be juxtaposed to the presumable dominance of annihilation topologies in the decays $D^+\to\pi^+\ell^+\ell^-$, as explained in Ref.~\cite{Bansal:2025hcf}. The so-called J-type topology, which is accounted-for in our analysis, corresponds to one type of the annihilation diagrams considered therein. Furthermore, in $D^+\to\pi^+\ell^+\ell^-$ the annihilation topologies are accompanied by the dominant Wilson coefficient $C_1$ and the emission topology comes with $C_2$, whereas in $D\to\pi\pi\ell^+\ell^-$ only the cascade decay comes with $C_1$; an analogous argument has been made in Ref.~\cite{Feldmann:2017izn}. In any case, the hierarchies pointed out in Ref.~\cite{Bansal:2025hcf} strictly apply to the unphysical region of $q^2$.
}
With regard to the cascade decay, an annihilation topology would come multiplied with the Wilson coefficient $C_2$, as opposed to the three times larger $C_1$ that multiplies the dominant emission topology.

We only consider the decay through $D^0\to a_1(1260)^+\pi^-$ and omit the CP-conjugate intermediate state $D^0\to a_1(1260)^-\pi^+$, which has different weak dynamics, as a naive estimate based on leading-order QCD factorization indicates a much smaller contribution to the branching ratio than the decay under study.
This finding is further supported by the quoted values for the decay width fractions in the amplitude analyses of $D^0\to4\pi$, of which the one of $D^0\to a_1(1260)^-\pi^+$ is about ten times smaller than that of $D^0\to a_1(1260)^+\pi^-$.

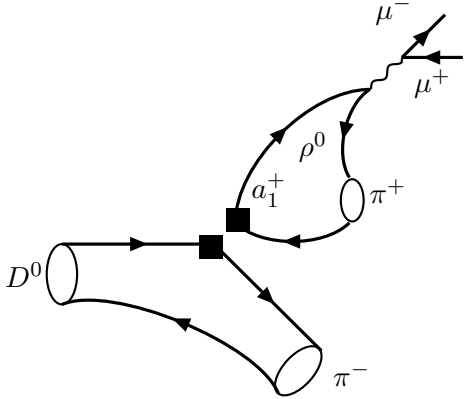
\begin{figure}[h!]
    \centering
    \resizebox{0.35\textwidth}{!}{
    \begin{tikzpicture}[thick]
        \begin{feynman}
            \vertex (a0);
            \vertex [below left=0.15cm of a0] (t1) {$D^0$};
            \vertex [right=2cm of a0] (a1);
            \vertex [below right=2cm of a1] (a2);
            \vertex [below right=0.02cm of a2] (t2) {$\pi^-$};
            \vertex [below=0.8cm of a0] (a3);
            \vertex [below left=0.8cm of a2] (a4);
            \vertex [above right=0.4cm of a1] (b0);
            \vertex [above right=0.1cm of b0] (t3) {$a_1^+$};
            \vertex [above right=2.5cm of b0] (b1);
            \vertex [below left=0.6cm of b1] (t4) {$\rho^0$};
            \vertex [right=1.5cm of b0] (b3);
            \vertex [above right=0.17cm of b3] (t2) {$\pi^+$};
            \vertex [above=0.6cm of b3] (b2);
            \vertex [above right=0.6cm of b1] (c1);
            \vertex [above right=0.8cm of c1] (c0);
            \vertex [right=0.8cm of c1] (c2);
            \diagram*{{(a0)--[fermion,very thick](a1)--[fermion,very thick](a2)},{(a4)--[quarter right,fermion,very thick,looseness=0.6](a3)},{(b3)--[quarter left,fermion,very thick,looseness=0.8](b0)--[quarter left,fermion,very thick,looseness=0.8](b1)--[quarter right,fermion,very thick,looseness=0.8](b2)},{(c2)--[fermion,very thick,edge label=$\mu^+$](c1)--[fermion,very thick,edge label=$\mu^-$](c0)},(b1)--[boson](c1)};
            \draw[] (0,-0.4) ellipse (0.2cm and 0.4cm);
            \draw[rotate around={-42:(3.11,-1.68)}] (3.11,-1.68) ellipse (0.2cm and 0.4cm);
            \draw[] (3.82,0.58) ellipse (0.15cm and 0.28cm);
            \node[shape=rectangle,fill=black] (n1) at (1.96, -0.04) {\rule{0.5mm}{0.5mm}};
            \node[shape=rectangle,fill=black] (n2) at (2.32, 0.32) {\rule{0.5mm}{0.5mm}};
        \end{feynman}
    \end{tikzpicture}
    }
    \caption{Cascade topology contributing to charm-meson $D^0$ decay: an intermediate $a_1^+$ resonance decays strongly to $ \rho^0 \pi^+ $, with subsequent $ \rho^0 \to \mu^+ \mu^- $ electromagnetic decay. The pair of filled squares corresponds to the effective four-quark weak interaction.}\label{fig:Cascade_Feynman}
\end{figure}

Under these assumptions
the only additional $S$-matrix element, with respect to previous works, contributing to the decay is 
\begin{eqnarray}
\label{eq:Smatrix-cascade}
    \langle\pi^+ \pi^- \ell^+ \ell^-| S |D^0\rangle^{(\mathrm{casc})}= 
    (2 \pi)^4 \, \delta^{(4)} (p + q - p_D) (\overline{u}_\ell\gamma_\mu v_\ell)\, \left( \lambda_d \frac{G_F}{\sqrt{2}} \,e^2  C_1(\mu)\right) \\
    \left( f_+(k^2)(p_1+2 p_2)^\mu +f_-(k^2)p_1^\mu\right) \, \frac{m_{a_1}f_{a_1} \, g_{a_1\rho\pi} }{P_{a_1}(k^2)}  
    \,
    \frac{1}{P_{\rho^0}(q^2)} 
    \,   \frac{f_{\rho^0}}{\sqrt{2}m_{\rho^0}}  
    \, B_{\mathrm{casc}}e^{i \delta_{\mathrm{casc}}}\,,  \nonumber
\end{eqnarray}
where $k^2$ is the squared momentum of the $a_1$ meson, $B_{\mathrm{casc}},\delta_{\mathrm{casc}}$ are the free normalisation factor and phase respectively, and $f_+(k^2), f_-(k^2)$ are the two $D\to\pi$ form factors.
The function $ P_{a_1}(k^2) $ is the Kuhn-Santamaria lineshape \cite{Kuhn:1990ad}, while $ P_{\rho^0}(q^2) $ is provided in our previous paper \cite{Fajfer:2023tkp}.
The insertion of the $Q_1^d$ operator under factorization leading to the cascade topology is illustrated in Figure~\ref{fig:Cascade_Feynman}.

As the cascade topology is kinematically different from the Q2B one, it manifests distinctly in the observables commonly presented in the experimental analyses, which are defined based on the kinematical variables $p^2 \equiv m(\pi^+ \pi^-)^2$, $ q^2 \equiv m(\ell^+ \ell^-)^2 $, $\theta_\pi$, $\theta_\ell$, $\phi$.\footnote{For a definition of the kinematical variables and angular observables see Ref.~\cite{Fajfer:2023tkp}.\label{fn:myfoot}} Namely, the squared momentum of the axial resonance takes the form 
\begin{equation}
    k^2=m_D^2+m_\pi^2-\frac{m_D^2+p^2-q^2}{2}-\frac{\sqrt{\lambda_h\lambda_D}}{2 p^2}\cos\theta_\pi \, ,
     \label{k2def}
\end{equation}
where $ \lambda_h = \lambda (p^2, m_\pi^2, m_\pi^2) $, $ \lambda_D = \lambda (m_D^2, p^2, q^2) $ are K\"all\'en functions.
While in the $q^2$ distribution a peak is still expected around the mass of the $\rho^0$ meson, from which the lepton pair is created, there is no sharp resonant peak associated with this topology in the $p^2$ distribution. On the other hand, in the distribution over $\cos\theta_\pi$ a smooth peak is expected around a value
determined by the kinematics, a feature which is absent in the purely Q2B-driven decays, wherein the shape of the distribution is parabolic. The exact shape of the distribution over $\cos\theta_\pi$ depends on the size of the cascade decay amplitude, as well as of the interferences with the Q2B topologies. Accordingly, all the angular observables discussed in Ref.~\cite{Fajfer:2023tkp} are expected to be modified to varying degrees. 

We calculate the amplitudes from the Q2B topologies as per our previous work \cite{Fajfer:2023tkp}. For the new cascade contribution, we implement the following: we use the value $C_1(m_c)=1.22$ at next-to-leading order (NLO) in the naive dimensional regularisation (NDR) scheme, and the $D\to\pi$ form factors  from the lattice \cite{FermilabLattice:2022gku}. The rest of the $a_1$-related constants appearing in the matrix element of Eq.~\eqref{eq:Smatrix-cascade}, namely $f_{a_1} \cdot g_{a_1\rho\pi}$, can be fitted along with $B_{\mathrm{casc}}$.

To facilitate the comparison between the effect on the observables of the new contribution and of the previously included contributions, we write the additional transversity form factors that the $a_1 \pi$ amplitude induces. The differential decay width can be expressed as a sum of angular observables as per Eqs.~(38)-(48) of Ref.~\cite{Fajfer:2023tkp}. Each angular observable $I_{1-9}$ results from a specific way of integrating the distribution over the angular variables $\theta_\ell$ and $\phi$. Further integration over the dipion angle $\theta_\pi$ in two different ways as indicated in Eq.~(54) of the same work results in a series of observables where the effect of each partial wave is distinct.
As  discussed therein, an approximate, effective Wilson coefficient $C_9^{\mathrm{eff:}S}$ and $C_9^{\mathrm{eff:}P}$ without the addition of the cascade contribution can be assigned separately to the amplitudes for which the pion pair is in an $S$- or a $P$-wave, multiplying the respective transversity form factors $\mathcal{F}_S$ and $\mathcal{F}_{P,\parallel,\perp}$. 

In the presence of the cascade topology discussed in this work, the angular observables $I_{1-9}$ can still be expressed as in the previous work \cite{Fajfer:2023tkp} but with the appropriate addition of all the contributions $\mathcal{F}_{casc}\cdot C_9^{\mathrm{eff:casc}}$, where the transversity form factors are the following:

\begin{equation}
    \mathcal{F}_{S,{\rm casc}}(p^2,q^2,\cos\theta_\pi)=-N g_{a_1\rho\pi}\frac{\sqrt{\beta_\ell(3-\beta_\ell^2)}\lambda_h^{1/4}\lambda_D^{3/4}}{2\sqrt{2}\sqrt{p^2}}\frac{1}{P_{a_1}(k^2)}\frac{3f_+(k^2)+f_-(k^2)}{2} \,,
\end{equation}

\begin{equation}
    \mathcal{F}_{\parallel ,{\rm casc}}(p^2,q^2,\cos\theta_\pi)=N g_{a_1\rho\pi}\frac{\sqrt{\beta_\ell(3-\beta_\ell^2)}\lambda_h^{3/4}\lambda_D^{1/4}\sqrt{q^2}}{\sqrt{2}p^2}\frac{1}{P_{a_1}(k^2)}\frac{-f_+(k^2)+f_-(k^2)}{2} \,,
\end{equation}

\begin{equation}
    \mathcal{F}_{P,{\rm casc}}(p^2,q^2,\cos\theta_\pi) = -\frac{(m_D^2-p^2-q^2)}{2 (p^2 \, q^2)^{1/2}} \mathcal{F}_{\parallel ,casc}(p^2,q^2,\cos\theta_\pi) \,,
\end{equation}
and $\mathcal{F}_{\perp,{\rm casc}} = 0$,
where the normalisation factor $N$ is the same one of Eq.~(52) of Ref.~\cite{Fajfer:2023tkp}, and the effective Wilson coefficient is

\begin{equation}
     C_9^{\mathrm{eff:casc}}(\mu; q^2)=8\pi^2 C_1(\mu) \frac{1}{P_{\rho^0}(q^2)} 
    \cdot \frac{m_{a_1}f_{a_1}f_{\rho^0}}{m_{\rho^0}} B_{\rm casc} e^{i\delta_{\rm casc}}\,.
\end{equation}

We note the most important implications that the presence of this additional dynamics has on the angular observables in the following:
firstly, it introduces the possibility of relative phases between transversity form factors of the same partial wave (e.g., induced by
the lineshapes of $a_1$ and $\rho^0$,
$1/P_{a_1}(k^2)$ and $1/P_{\rho^0} (p^2)$
in $\mathcal{F}_P$ and $\mathcal{F}_\parallel$).
Consequently, observables that require such a non-vanishing relative phase can now obtain non-zero values: this is the case for some of the LHCb-measured observables \cite{LHCb:2021yxk}, both the SM-dominated ones $S_8\equiv\langle I_8 \rangle_-$ and $S_9\equiv\langle I_9\rangle_+$ as well as the null test $S_7\equiv\langle I_7\rangle_-$.\hyperref[fn:myfoot]{\textsuperscript{\getrefnumber{fn:myfoot}}}

Another difference that the cascade topology introduces with respect to the purely Q2B decay is the explicit dependence of the transversity form factors on the angle $\theta_\pi$, through the dependence of the $a_1$ lineshape and of the $D\to\pi$ form factors on the momentum $k^2$ of Eq.~\eqref{k2def}. This non-trivial dependence means in turn that the $\theta_\pi$-integrated observables $\langle I_i\rangle_{\pm}$ can
no longer be strictly expressed in terms of separate $S$-only, $P$-only or $S$-$P$ interference components as per Eqs.~(55)-(72) of Ref.~\cite{Fajfer:2023tkp}, if the contribution of the cascade topology is sizable. As a result, the observables $\langle I_3\rangle_- $ and $\langle I_9\rangle_-$, as well as the null test $\langle I_6 \rangle_-$, all previously thought to vanish irrespective of the hadronic model used and of the presence of NP, can now obtain non-zero values. $\langle I_6 \rangle_-$ can then be utilised as an additional observable for the discovery of NP effects from its interference with the cascade SM component, and $\langle I_3\rangle_-, \, \langle I_9\rangle_-$ are apt observables for a clear signal of the existence of the cascade topology, if one selects appropriate integration limits for $p^2$ and $q^2$.

\section{Fitting to LHCb data and predictions}\label{sec:fitting}

\begin{figure}[t]
    \centering
    \includegraphics[width=0.5\linewidth]{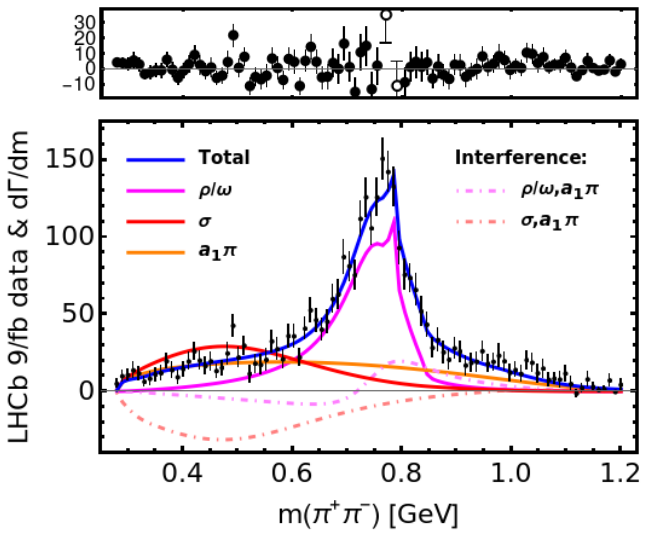}
    \caption{Fit to the differential branching ratio as a function of the invariant mass of the pion pair, compared to LHCb binned data. The best fit curve, together with the main individual contributions, are displayed. The solid blue line represents the full contribution, while solid magenta, red and orange represent the individual $\rho^0/\omega \to \pi^+ \pi^-$, $\sigma \to \pi^+ \pi^-$ and cascade contributions, respectively. Dot-dashed curves represent interference terms with the cascade contribution.
    The empty data points pertain to the sharp $\omega$ peak region.
    }
    \label{fig:p2_distribution}
\end{figure}

\begin{figure}[t]
    \centering
    \includegraphics[width=0.5\linewidth]{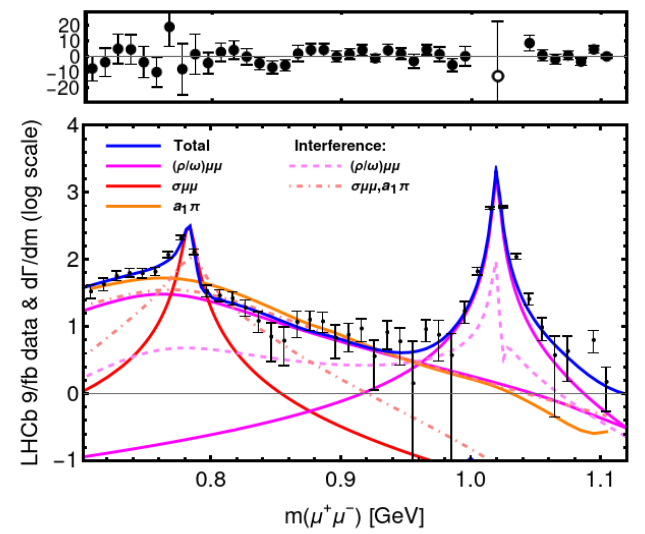}
    \caption{
    Similar to Figure~\ref{fig:p2_distribution}, where now the invariant mass of the muon pair is considered.
    The dot-dashed curves represent the (absolute values of the) interference terms of the $\sigma$ with the cascade contributions, while the dashed curve represents the (absolute value of the) interference of the $ \rho^0 \rho^0 $ and $ \rho^0 \phi $ contributions.
    The empty data point pertains to the sharp $\phi$ peak region.
    }
    \label{fig:q2_distribution}
\end{figure}

We now discuss the comparison of our model to the binned differential branching ratios of Ref.~\cite{LHCb:2021yxk}.
Figures~\ref{fig:p2_distribution} and \ref{fig:q2_distribution} display the comparison between LHCb data in bins of $q^2$ and $p^2$, respectively, and our theoretical modeling, after adjusting its free parameters from the fit.

Some precision is in order. Starting with the data in use, we consider $p^2$ values ranging from the threshold $4 m_\pi^2$ up to $0.18$~GeV$^2$, and from $0.32$~GeV$^2$ up to $1.0$~GeV$^2$, extending the range of our previous analysis \cite{Fajfer:2023tkp}.
The region $[0.18, 0.32]$~GeV$^2$ is excluded from our fit since there is likely contamination from a different charm-meson decay mode having a $K_S^0$ in the final state.
Beyond $p^2 = 1.0$~GeV$^2$, features such as heavier resonances that are less known compared to the lightest ones included in the analysis start playing a role.
Accordingly, further small contributions are needed in the high-energy region above $p^2 = 1.0$~GeV$^2$ as seen from Figure~\ref{fig:p2_distribution}
(since data points in this region tend to be distributed above the theoretical curve), which however is not included in our fit.

Since the $\omega \to \pi^+ \pi^-$ contribution represents a sharp peak, we collect the two bins above and below its nominal mass into two wider bins, in order to circumvent resolution effects as discussed in the case of $\phi \to \mu^+ \mu^-$ in Ref.~\cite{Fajfer:2023tkp}. As therein, we neglect correlations among bins, which are not reported by LHCb. The two resulting broader bins are shown as empty circles in Figure~\ref{fig:p2_distribution}
(while the empty circle in Figure~\ref{fig:q2_distribution}
corresponds to the $\phi \to \mu^+ \mu^-$ wider bin). Since this isospin-violating contribution is anyways very small, to simplify the global fit we fix its relative size and phase with respect to $\rho^0 \to \pi^+ \pi^-$ to the values obtained in our previous work \cite{Fajfer:2023tkp}.\footnote{The solid curves in Figure~\ref{fig:p2_distribution}
are built from connecting the predictions for each final $p^2$ bin, and therefore the $\omega$ contribution does not appear as a sharp peak.}

Similarly,
we take the lowest value of $q^2$ at $0.5$~GeV$^2$ in order to avoid effects not included in our analysis such as heavier resonant states manifesting at high $p^2$, and the highest value of $q^2$ at $1.2$~GeV$^2$, the endpoint of available binned data.
In total, there are 94 experimental bins in use.
The resulting $(p^2, q^2)$ region we analyse is populated with lighter resonances which are relatively well known.

The addition of the cascade contribution improves the fit substantially.
The improvement comes from the better description of the $p^2$ differential branching ratio.
Indeed, the cascade contribution has a long tail into large $p^2$ values (longer than displayed by the $\sigma \to \pi^+ \pi^-$ profile), and its inclusion improves in particular the description of the region above the $\rho^0 \to \pi^+ \pi^-$ resonance; the descriptions of the region below this resonance, and the region dominated by it, also improve.
We obtain a $ \chi^2_{\rm min} $ of $82$, or a $p$-value of about $48\%$ ($N_{\rm d.o.f.} \simeq 82$), thus drastically improving our past analysis where a $ \chi^2_{\rm min} / N_{\rm d.o.f.} $ of about $2$ was found \cite{Fajfer:2023tkp}. Since theoretical uncertainties from the use of large-$N_C$ counting are very difficult to estimate reliably, only statistical uncertainties are included in the $\chi^2$.
This is so because estimating such uncertainties requires some knowledge about beyond-the-leading-order contributions in the large-$N_C$ counting, which is beyond the scope of the present analysis. Nonetheless, part of their effects is included in the free normalization factors.

A slightly better fit is obtained at a configuration where the parameter $a_S(0)$, setting the overall size of the $\sigma$ contribution, is considerably larger, which however is in tension with the analogous parameter extracted from the semileptonic $ D^+ \to \pi^- \pi^+ \ell^+ \nu_\ell$ transition \cite{BESIII:2018qmf}.
We focus on the configuration presented in the following since it is more consistent with large-$N_C$ counting, while the other solution can be found in Appendix~\ref{app:alternative_solution}.
We note however that a contribution from the annihilation topology to the rare and non-leptonic decays, which is absent in semileptonic decays, could be absorbed into $\sigma$ normalization factors.

Note that $\sigma \to \pi^+ \pi^-$ and $\rho^0 \to \pi^+ \pi^-$, which are respectively in $S$- and $P$-waves, do not interfere between them, while the cascade contribution interferes with both. This latter fact allows increasing the sizes of the $\sigma$ and cascade amplitudes for a better fitting, while adding a destructive interference in the low-$p^2$ region.
Indeed, there is a sizable cancellation in $p^2$ between the $\sigma$-only contribution and the interference term of the $\sigma$ contribution
with the cascade.
In Figure~\ref{fig:p2_distribution} we also present the residual difference between LHCb data and our model evaluated at the best fit point.
Figure~\ref{fig:q2_distribution} shows the agreement of LHCb data in bins of $q^2$ and our theoretical modeling.

The main components of the fit are $ \rho^0 \phi $, $ \rho^0 \rho^0 $, $ \sigma \omega $, $ \sigma \phi $, $ a_1^+ \pi^- $, and interference terms among the cascade and $\sigma \to \pi^+ \pi^-$ and $\rho^0 \to \pi^+ \pi^-$ contributions
(when two resonances in Q2B topologies are shown, the first decays to the pion pair, and the second to the lepton pair).
As discussed in Ref.~\cite{Fajfer:2023tkp}, the $ \sigma \omega $ contribution is needed to describe the $ \omega \to \mu^+ \mu^- $ peak.
Indeed, there is a cancellation between two topologies (namely, of the so-called W- and J-types) when assessing the contribution from $\rho^0 \omega$ in large-$N_C$.
Moreover, $ a_1^+ \to \omega \pi^+ $ is suppressed.
The fit fractions are about 60\% for the $\rho^0$, and about 40\% for both the $\sigma$ and the cascade contributions. Interference terms among the latter two categories produce a destructive contribution of about 40\%.
We remind the reader that the size of the $S$-wave contribution in semileptonic decays $D^+\to\pi^-\pi^+ e^+\nu$ \cite{BESIII:2018qmf} is also large, of about $25\%$.

To constrain the free parameters, it has proven fundamental to combine the two differential distributions. For instance, not including the cascade contribution can give a suitable description of the $d \Gamma / d p^2$ data by enhancing the $\rho^0 \rho^0$ contribution while suppressing the size of $\rho^0 \phi$, but this does not lead to a satisfactory description of the $d \Gamma / d q^2$ data.
The $q^2$ distribution plays a pivotal role in fixing the $\rho^0 \phi$, $\sigma \omega$, and $ a_1 \pi $ contributions, while the $p^2$ distribution provides detailed information about the line-shapes manifesting in the $\pi^+ \pi^-$ invariant mass, since it can distinguish resonances paired with the $\phi \to \mu^+ \mu^-$ from those paired with $ \rho^0 $ or $\omega \to \mu^+ \mu^-$ due to the difference in the allowed $(p^2, q^2)$ phase space.

We fix the parameter $ B^{(S)}_{\rho^0} = B^{(P)}_{\rho^0} = B_{\rho^0} $, since the value of $a_S (0)$ is left free in the fit. We extract the following ranges

\begin{eqnarray}
    && A_1 (0) B_{\rho^0} = 1.2 \pm 0.1 \,, \\
    && B_\phi / B_{\rho^0} = 0.79 \pm 0.04 \,, \\
    && B^{(S)}_\omega / B^{(S)}_{\rho^0} = 1.3 \pm 0.1 \,, \\
    && B^{(S)}_\phi / B^{(S)}_{\rho^0} = 0.48 \pm 0.04 \,, \\
    && a_S(0) / A_1 (0) = ( 48 \pm 3 ) \; \text{GeV} \,, \\
    && g_{a_1 \rho \pi} f_{a_1} B_{\rm casc} = ( 3.6 \pm 0.1 ) \; {\rm GeV}^2
    \,,
\end{eqnarray}
given at $1\sigma$ henceforth.
The fit finds
some negative correlations among the normalization factors, which thus avoids having too many additive contributions to the differential branching ratio: in particular, between the two parameters controlling the $\phi$ peak ($ B_\phi / B_{\rho^0} $ and $ B^{(S)}_\phi / B^{(S)}_{\rho^0} $, between which a linear correlation of $ \rho = -0.7 $ is obtained), and between the parameter $ a_S(0) / A_1 (0) $ setting the overall normalization of the $\sigma$ contribution and the two other normalization factors related to this contribution (namely, $ B^{(S)}_\omega / B^{(S)}_{\rho^0} $ and $ B^{(S)}_\phi / B^{(S)}_{\rho^0} $, with linear correlations of $ \rho = -0.8 $ and $ \rho = -0.7 $, respectively).
The value of $ A_1 (0) B_{\rho^0} $ is extracted from Ref.~\cite{LHCb:2017uns}.
These fit results are more consistent with the large-$N_C$ counting than in our previous Ref.~\cite{Fajfer:2023tkp}, since in particular $B^{(S)}_\phi / B^{(S)}_{\rho^0}$ is significantly larger than before (cf. Appendix~\ref{app:alternative_solution} below).\footnote{The extracted value of $a_S(0)$ was incorrectly reported in Ref.~\cite{Fajfer:2023tkp}, the correct range being $ 48 \; \text{GeV} \lesssim a_S(0) / A_1 (0) \lesssim 75 \; \text{GeV} $, i.e., a factor of $1.2$ larger.}

The interference terms among amplitudes allow the extraction of relative phases, by the analysis of the differential branching ratios as a function of $q^2$ and of $p^2$.
The addition of the cascade contribution, which can interfere with both $S$- and $P$-waves adds two new relative strong phases in the fit.
The fit is sensitive to the following phases

\begin{eqnarray}
    && \delta_{\{\sigma , \rho^0\}} - \delta_{\{\sigma , \omega\}} = (-0.57 \pm 0.03) \pi \,, \\
    && \delta_{\{\rho^0 / \omega , \rho^0\}} - \delta_{\{\rho^0 / \omega , \phi\}} = (-0.53 \pm 0.09) \pi \,, \\
    && \delta_{\rm casc} - \delta_{\{\rho^0 / \omega, \rho^0\}} = (0.9 \pm 0.2) \pi \,, \\
    && \delta_{\rm casc} - \delta_{\{\sigma, \rho^0\}} = (0.50 \pm 0.02) \pi \,,
\end{eqnarray}
while $ \delta_{\{\sigma , \rho^0\}} - \delta_{\{\sigma , \phi\}} $ is poorly determined.
A different, very similar best fit configuration is found by flipping the sign of $ \delta_{\{\rho^0 / \omega , \rho^0\}} - \delta_{\{\rho^0 / \omega , \phi\}} $.
Inspecting the values above, we find phase differences compatible with values as large as $ \pm \pi /2 $.
These large phases should represent an important difficulty if attempting to calculate them from perturbative techniques.
The presence of the cascade contribution leverages the cascade-Q2B interference terms, which are present in the mass distributions, for the determination of the difference of phases $ \delta_{\{\rho^0 / \omega, \rho^0\}} - \delta_{\{\sigma, \rho^0\}} $, which is otherwise currently not accessible.
The determination of the relative strong phases is important in looking for NP: for instance, much above the $\phi$ resonance where the SM contribution is suppressed, but where it is subjected to large uncertainties due to the interference terms \cite{DeBoer:2018pdx}.

\begin{figure}[t]
    \centering
    \includegraphics[width=0.5\linewidth]{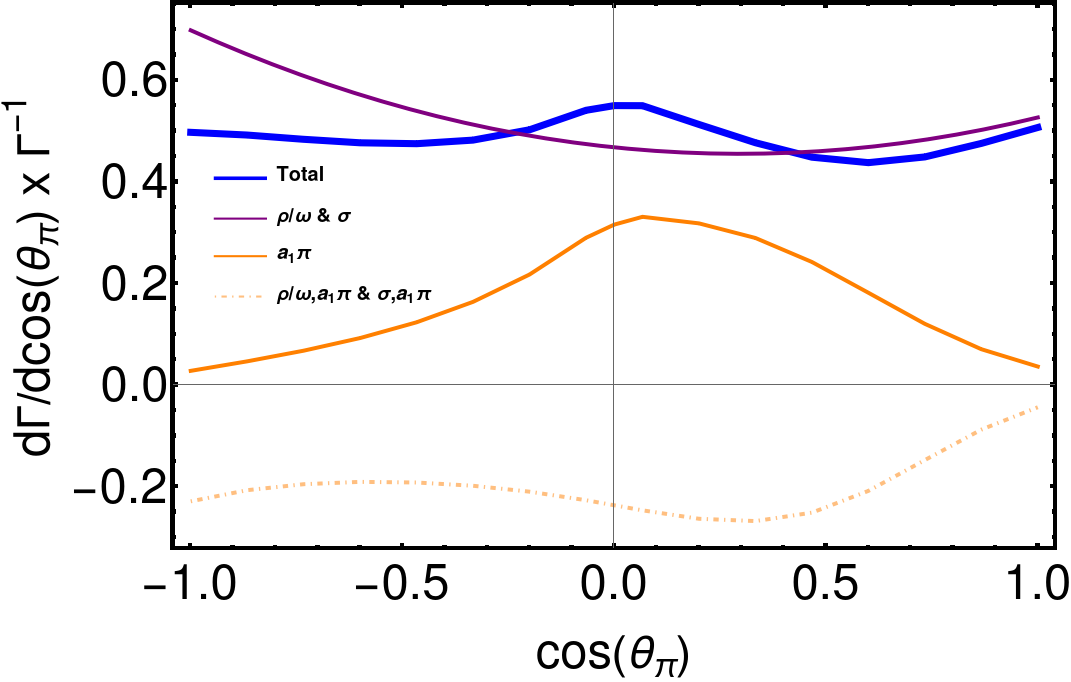}
    \caption{The differential decay rate as a function of $\cos ( \theta_\pi )$, with parameters at their best-fit values from the fit to the mass distributions of Figures~\ref{fig:p2_distribution} and \ref{fig:q2_distribution}. The thinner continuous lines correspond to the Q2B-only (purple) and cascade-only (orange) distributions, while the dash-dot line corresponds to the Q2B-cascade interference. In thick blue we show the total distribution, for which a distinctive bump is attributed to the cascade component.}
    \label{fig:dGammadthetapi}
\end{figure}

Having fixed the parameters from the differential mass distributions, we now shift to predictions for other observables.
First of all, we can provide a prediction for the distribution $d\Gamma/ d\cos\theta_\pi$, which as previously discussed in Ref.~\cite{Fajfer:2023tkp} also probes the $S$-wave contribution; see Figure~\ref{fig:dGammadthetapi}. A measurement of this distribution would thus further validate both the cascade contribution and the $\sigma$-only component of the decay and help in probing the model parameters, especially relative phases.
Concerning the angular observables, in light of the large uncertainties both of the currently available experimental angular analysis as well as of the theoretical prediction of those observables, we refrain from performing
a comprehensive comparison of the measured observables $S_2 \equiv \langle I_2 \rangle_+$, $S_3 \equiv \langle I_3 \rangle_+$ and $S_4 \equiv \langle I_4\rangle_-$,\hyperref[fn:myfoot]{\textsuperscript{\getrefnumber{fn:myfoot}}} which are driven by the same ingredients as the differential distributions and presented a good agreement between prediction and experiment already in our previous analysis \cite{Fajfer:2023tkp}. Here we focus on providing estimates for the observables that obtain non-vanishing values only due to the cascade topology.

We find that the SM observables $\langle I_8\rangle_-$ and $\langle I_9\rangle_+$ obtain values of a few percent in the dimuon-mass bins around the $\rho^0$ peak. The large value of $\langle I_9\rangle_+$ measured in LHCb with more than a 3$\sigma$ significance in the bin on the left of the $\phi$ peak is not reproduced in our analysis, as the cascade topology only proceeds through an emission of a $\rho^0(\to\ell^+\ell^-)$, thus interfering less prominently with the Q2B topology $D^0 \to \rho^0 (\to\pi^+\pi^-)\phi(\to\ell^+\ell^-)$.  The null test $\langle I_7\rangle_-$ as defined by LHCb is estimated to give less than $1\%$ when saturating the current bounds for NP-induced $C_{10}$ \cite{Fajfer:2015mia, Fuentes-Martin:2020lea}. However, if one limits the integration over $p^2$ and $\cos\theta_\pi$ in the region where there is maximal interference between the cascade topology and $D^0\to\rho^0\ell^+\ell^-$ (namely, for $p^2$ either only left or only right of the $\rho^0$ peak), we stress that a potential signal of NP can be enhanced in this observable. 
Finally, the angular observables $\langle I_3\rangle_-$, $\langle I_6\rangle_-$ and $\langle I_9\rangle_-$, which can have non-zero values because of the non-trivial angular dependence of the cascade amplitude, are also found to give negligible values compared to the near-future experimental sensitivity when integrated as per the LHCb-defined $q^2$ bins, but as before this picture would change for optimized observables.

\section{Comparison to non-leptonic decays}\label{seb:comparisons}

For a direct comparison of the rare decays under study to the dynamics of the non-leptonic (NL) decays, we assume the same framework of weak decays to the intermediate mesons complemented by resonant lineshapes and a complex coefficient. Accordingly, we write down the decay amplitudes as 

\begin{eqnarray}
     && \langle \pi^+(p_1)\pi^-(p_2)\pi^+(q_1)\pi^-(q_2) |S |D^0\rangle^{\mathrm{(casc,NL)}} = 
      (2 \pi)^4 \, \delta^{(4)} (p + q - p_D) \, \left( \lambda_d \frac{G_F}{\sqrt{2}}  C_1(\mu)\right) \nonumber \\
  && \left( f_+(k^2)(p_1+2 p_2)^\mu +f_-(k^2)p_1^\mu\right) \frac{m_{a_1}f_{a_1} \, g_{a_1\rho\pi} }{P_{a_1}(k^2)}
    \cdot\frac{1}{P_\rho(q^2)} 
    \cdot b_\rho (q_{1\mu}-q_{2\mu}) 
    \cdot B_{\mathrm{casc, NL}}\, e^{i\delta_{\mathrm{casc, NL}}} \nonumber \\
\end{eqnarray}
for the $a_1$-mediated amplitude, and similarly for the other decay modes through $\rho\rho,\rho\phi$, $\sigma\rho,\sigma\phi$, with different complex coefficients in each case.

In contrast to the rare decays, where various subtleties have to be accounted for in different kinematical regions given the overall small branching ratio (e.g., Bremsstrahlung in the low $q^2$ or short-distance contributions away from the resonances), the non-leptonic decays are assumed to be well described by the sum of all resonance-mediated partial amplitudes in the whole available phase space. Therefore, we extract the free normalization and phase parameters from the reported integrated branching fractions of each partial amplitude (called fit fractions in the amplitude analyses) or the integrated branching fractions resulting from the interference of two different partial amplitudes (called interference fractions in the amplitude analyses).\footnote{
We note that
Ref.~\cite{Bansal:2025hcf} uses non-leptonic three-body decays for a similar exercise.} The values of the free normalization factors extracted when using the results of Ref.~\cite{dArgent:2017gzv} are presented on Table \ref{tab:comparison}. We do not show the phase differences that can be probed from the non-leptonic decays. Those are only the ones of the interferences of the cascade topology with the $\rho\rho$ as well as with the $\sigma\rho$, as all the other pairs of topologies interfering in the rare decays appear instead non-simultaneously in the $4\pi$ or in the $2\pi 2 K$ modes. The fitted phase differences differ substantially between the rare and the non-leptonic decays.

In the case of the Q2B topologies to two vector mesons $D\to\rho\rho$ and $D\to\rho\phi$, we compare our calculation with the sum of the three waves $S, P, D$ in the experimental work. In our description of large-$N_C$ corrected by a complex coefficient commonly assigned to the whole intermediate state, $B_{\rho\rho,\mathrm{NL}}e^{i\delta_{\rho\rho,\mathrm{NL}}}$ and $B_{\rho\phi,\mathrm{NL}}e^{i\delta_{\rho\phi,\mathrm{NL}}}$, the higher partial waves are calculated to have increasingly smaller contributions when using the values of the form factors extracted from the semileptonic decays $D\to\pi\pi e\nu$ \cite{BESIII:2018qmf}. We note that both amplitude analyses \cite{dArgent:2017gzv,BESIII:amplitudean} observe instead an inverse hierarchy of the partial waves, where the $D$-wave dominates over the $P$-wave, which dominates over the $S$-wave. This apparent departure from large-$N_C$ lies beyond the scope of our current work.

\begin{table}
\resizebox{\textwidth}{!}{
\begin{tabular}{c|c||c||c }
Contribution & Parameter [units] & Non-leptonic decays & Rare decays  \\ \hline
cascade & $g_{a_1\rho\pi }\, f_{a_1}\, B_\mathrm{casc}$ [GeV$^2$]  &  $3.5 \pm 0.3 $ & $3.6\pm 0.1$ \\
S-wave $\sigma\rho$ &  
$a_S(0) B_{\rho^0}^{(S)}$ [GeV] & 
$70 ^{+20}_{-40} $ 
& $58 \pm 6$  \\
S-wave $\sigma\phi$ & $a_S(0)B_\phi^{(S)}$ [GeV] & 
$20^{+8}_{-14}$ & $28 \pm 3$  \\
P-wave $\rho\rho$ & $A_1(0)B_{\rho^0}^{(P)}$ [1] &  $1.5\pm 0.3 $
& $1.2\pm 0.1 $  \\
P-wave $\rho\phi$ & $A_1(0)B_\phi^{(P)}$ [1] &  $1.04 \pm 0.08 $ 
& $0.95\pm 0.09$  \\ 
\hline 
\end{tabular} 
}
\caption{Comparison of the $1\sigma$ ranges for the normalization parameters extracted from the amplitude analysis of Ref.~\cite{dArgent:2017gzv} of the non-leptonic decays $D^0\to\pi^+\pi^-\pi^+\pi^-$ and $D^0\to\pi^+\pi^- K^+K^-$ and from our nominal fit to the rare-decay $D^0\to \pi^+\pi^-\mu^+\mu^-$ data.}
\label{tab:comparison}
\end{table}

Remarkably, we find an excellent agreement between rare and the non-leptonic decays $D^0\to \pi^+\pi^-\pi^+\pi^-$ and $D^0\to \pi^+\pi^- K^+K^-$ for the normalization factor of the cascade topology, which is the partial amplitude measured with the smallest relative uncertainty among the modes that are common in non-leptonic and rare decays. The fit fraction of the $a_1\pi$ intermediate topology is also consistent between the CLEO-c \cite{dArgent:2017gzv} and the BESIII \cite{BESIII:amplitudean} analyses and largely dominant in both, although the latter does not separate the contribution of $a_1\to\sigma\pi$ (not relevant for the rare decays) from $a_1\to\rho\pi$. Further comparison of the $a_1$ parameters to direct strong-decay data point to a large corrective factor related to charm decays, as explained in Appendix \ref{section:a1}.

The rest of the normalization factors also exhibit a good agreement both between the rare decays and the amplitude analyses, and among the three amplitude analyses \cite{dArgent:2017gzv, LHCb:2018mzv, BESIII:amplitudean}, albeit with large uncertainties in all analyses.

\section{Conclusions}\label{sec:conclusions}

In this work we include for the first time in rare charm decay analyses the $a_1\pi$ partial amplitude. This feature results in a much improved description of the available experimental data, reflected most prominently on the invariant-pion-mass spectrum, and in a series of angular observables obtaining non-zero values. The novel cascade-topology component is found to contribute at the same level as the common two-vector-meson modes, in line with the findings of the amplitude analyses of four-body non-leptonic decays. Instead of attempting to isolate regions of the phase space that are clean from SM background, the existence of various sizable SM contributions manifesting with different angular structures gives the opportunity to constrain NP from a multitude of fairly independent, interference-induced null-test observables.
We believe that the current work is a valuable step in this endeavour. With the next experimental analyses, provided that they include a comprehensive set of observables, such as the five-fold differential distribution or optimally integrated angular observables (including the ones advocated for in Ref.~\cite{Fajfer:2023tkp}), the $d\Gamma/ d\cos\theta_\pi$ distribution and the $\pi^+\mu^+\mu^-$ spectrum, additional information can be utilised to test the required ingredients of hadronic origin and subsequently competitive bounds on NP can be set. More generally, this analysis advances the effort to better understand QCD dynamics in charm decays, which is crucial for determining the origin of the yet unexplained CP violation observed. In the future we plan to extend our analysis of $D^0\to \pi^+\pi^-\ell^+\ell^-$ to a combined one with $D^0\to K^+K^-\ell^+\ell^-$ as well as radiative decays.

\section*{Acknowledgements}

We would like to thank Svjetlana Fajfer for initial involvement in this project.
We are grateful to Anshika Bansal, Gudrun Hiller, Gino Isidori, Luka Leskovec, Dominik S. Mitzel, Tommaso Pajero, Antonio Pich, Sasa Prelovsek Komelj, Pablo Roig, and Dominik Suelmann for various discussions.
E.S. thanks Alex Lenz for the hospitality in the University of Siegen during the months in which part of this project was developed, and for providing financial support through the Deutsche Forschungsgemeinschaft (DFG, German Research Foundation) grant 396021762 - TRR 257.
E.S. gratefully acknowledges financial support from the Generalitat Valenciana (Spain) through the GenT program (CIDEGENT/2021/037), from the DAAD organization through the program Research Grants in Germany, 2025 (57742125) and from the Swiss National Science Foundation (Project No. 10003620). 
L.V.S. is supported by the Spanish Government (Agencia Estatal de Investigaci\'{o}n MCIN/AEI/10.13039/ 501100011033) Grants No. PID2020–114473GB-I00 and No. PID2023-146220NB-I00, and CEX2023-001292-S (Agencia Estatal de Investigaci\'{o}n MCIU/AEI (Spain) under grant IFIC Centro de Excelencia Severo Ochoa).

\appendix

\section{Alternative solution}\label{app:alternative_solution}

For the alternative best fit point mentioned in the main text,
we obtain a $ \chi^2_{\rm min} $ of $79$, or a $p$-value of about $58\%$ ($N_{\rm d.o.f.} \simeq 82$).
In this case,
the fit fractions are about 60\% for the $P$-wave, and about 60\% and 50\% for the $S$-wave and the cascade contributions, respectively. Interference terms among the latter two categories produce a destructive contribution of about 70\%.

We obtain the following normalization factors

\begin{eqnarray}
    && A_1 (0) B_{\rho^0} = 1.0 \pm 0.1 \,, \\
    && B_\phi / B_{\rho^0} = 0.92 \pm 0.07 \,, \\
    && B^{(S)}_\omega / B^{(S)}_{\rho^0} = 0.59 \pm 0.03 \,, \\
    && B^{(S)}_\phi / B^{(S)}_{\rho^0} = 0.22 \pm 0.02 \,, \\
    && a_S(0) / A_1 (0) = ( 90 \pm 6 ) \; \text{GeV} \,, \\
    && g_{a_1 \rho \pi} f_{a_1} B_{\rm casc} = ( 4.1 \pm 0.2 ) \; {\rm GeV}^2
    \,.
\end{eqnarray}
As advertised in the main text, $ a_S(0) / A_1 (0) $ is now substantially larger.
As in Ref.~\cite{Fajfer:2023tkp}, a suppressed value of $B^{(S)}_\phi / B^{(S)}_{\rho^0}$ is found. The same pattern of negative linear correlations is also observed here.
The extracted ranges for the relative angles are

\begin{eqnarray}
    && \delta_{\{\sigma , \rho^0\}} - \delta_{\{\sigma , \omega\}} = (-0.85 \pm 0.03) \pi \,, \\
    && \delta_{\{\rho^0 / \omega , \rho^0\}} - \delta_{\{\rho^0 / \omega , \phi\}} = (-0.40 \pm 0.05) \pi \,, \\
    && \delta_{\rm casc} - \delta_{\{\rho^0 / \omega, \rho^0\}} = (0.9 \pm 0.2) \pi \,, \\
    && \delta_{\rm casc} - \delta_{\{\sigma, \rho^0\}} = (0.76 \pm 0.01) \pi \,.
\end{eqnarray}
As before, a different very similar best fit configuration is also found by flipping the sign of $ \delta_{\{\rho^0 / \omega , \rho^0\}} - \delta_{\{\rho^0 / \omega , \phi\}} $.

\section{Comparison of $a_1$ parameters to other processes}\label{section:a1}

The normalization factor related to the $a_1$ resonance can be compared to estimates extracted from other environments. Because of our simple modelling for the transition $a_1\to \rho\pi$ with Eq.~\eqref{eq:a1_rho_pi_ME}, comparisons with more sophisticated analyses, e.g., within resonance chiral theory, are difficult. However, taking into account the on-shell estimate from Ref.~\cite{Zanke:2021wiq}, $ g_{a_1\rho\pi}\in [3.7,5.7] $~GeV,
and the value for the decay constant from Ref.~\cite{Yang:2007zt}, $f_{a_1}$=0.238 GeV, we find that $g_{a_1\rho\pi} \cdot f_{a_1} \in [0.88,1.36]$~GeV$^2$. A comparison to both the non-leptonic and the rare decays (first line of Table~\ref{tab:comparison}) indicates a large corrective normalization factor $B_\mathrm{casc}$ of around 3 for both classes of decays, which coincidentally is in reasonable agreement with our extracted value for $B_{\rho^0}^{(P)}$ when taking $A_1(0)=0.36$ \cite{BESIII:2018qmf} (fourth line of Table~\ref{tab:comparison}). In any case, we reinstate that with the implemented model we do not claim to be able to capture the detailed dynamics of the $a_1$ hadronic decays.

\bibliography{bib-cascade}{}
\bibliographystyle{unsrturl}

\end{document}